\documentstyle[fleqn,epsfig]{article}
\newcommand{\qsq    }{\mbox{$Q^2$}}
\newcommand{\psq    }{\mbox{$P^2$}}
\newcommand{\qzm    }{\mbox{$\langle \qsq \rangle$}}
\newcommand{\pzm    }{\mbox{$\langle \psq \rangle$}}
\newcommand{\gev    }{\mbox{$\rm GeV$}}
\newcommand{\gevsq  }{\mbox{$\rm GeV^2$}}
\newcommand{\der    }{\mbox{${\mathrm d}$}}
\newcommand{\dsigdx }{\mbox{$\der\sigma/\der x$}}
\newcommand{\ftqed  }{\mbox{$F_{\mathrm{2}}^{\gamma}$}}
\newcommand{\faqed  }{\mbox{$F_{\mathrm{A}}^{\gamma}$}}
\newcommand{\fbqed  }{\mbox{$F_{\mathrm{B}}^{\gamma}$}}
\newcommand{\ft     }{\mbox{$F_{2,{\rm had}}^{\gamma}$}}
\newcommand{\stt    }{\mbox{$\sigma_{\mathrm{TT}}$}}
\newcommand{\slt    }{\mbox{$\sigma_{\mathrm{LT}}$}}
\newcommand{\stl    }{\mbox{$\sigma_{\mathrm{TL}}$}}
\newcommand{\sll    }{\mbox{$\sigma_{\mathrm{LL}}$}}
\newcommand{\ttt    }{\mbox{$\tau_{\mathrm TT}$}}
\newcommand{\ttl    }{\mbox{$\tau_{\mathrm TL}$}}
\newcommand{\barph  }{\mbox{$\bar{\phi}$}}
\newcommand{\cosph  }{\mbox{$\cos\barph$}}
\newcommand{\costph }{\mbox{$\cos 2\barph$}}
\newcommand{\mumu   }{\mbox{$\mu\mu$}}
\newcommand{\az     }{\mbox{$\chi$}}
\newcommand{\aem    }{\mbox{$\alpha$}}
\newcommand{\faoft  }{\mbox{$\faqed/\ftqed$}}
\newcommand{\fboft  }{\mbox{$\fbqed/\ftqed$}}
\newcommand{\chidof }{\mbox{$\chi^2/\mathrm{dof}$}}
\newcommand{\invpb  }{\mbox{$\mathrm{pb}^{-1}$}}
\newcommand{\Y      }[3]{\mbox{$#1\,^{+\,#2}_{-\,#3}$}}

\newcommand{\omfwq  }{\mbox{${\mathcal{O}}({m_\mu^2}/{W^2})$}}
\begin{document}
\begin{center}
 {\huge\bf The QED Structure of the Photon}\\\vspace{0.4cm}
 Richard Nisius, on behalf of the LEP 
 Collaborations\footnote{Invited talk given at the 7th International 
 Workshop on Deep Inelastic Scattering and QCD, April 19 to 23, 1999,
 Zeuthen, to appear in the proceedings}\\
 \it{CERN, CH-1211 Gen\`eve 23, Switzerland}
\end{center}
%
%
%
\begin{abstract}
 Measurements of the QED structure of the photon based on the reaction 
 $\mathrm{e}\mathrm{e} \rightarrow \mathrm{e}\mathrm{e}
 \gamma^{(\star)}(\psq)\gamma^{\star}(\qsq)
 \rightarrow \mathrm{e}\mathrm{e} \mathrm{\mu}\mathrm{\mu}$
 are discussed. 
 This review is an update of the discussion of the results on the QED 
 structure of the photon presented in Refs.~\cite{NISIUS}, and 
 covers the published measurements of the photon structure functions
 \ftqed, \faqed\ and \fbqed\ and of the differential 
 cross-section \dsigdx\ for the exchange of two virtual photons.
\end{abstract}
%
%
%
%
\section{Introduction}
 Several measurements concerning the QED structure of the 
 photon have been performed by various experiments.
 Prior to LEP, mainly the structure function \ftqed\ 
 of the quasi-real photon, $\gamma$, as a function
 of the photon virtuality \qsq\ of the virtual photon, $\gamma^{\star}$,
 was measured.
 The LEP experiments refined the analysis of the \mumu\ final state, 
 and derived more information on the QED structure of the photon.
 The interest in the investigation of the QED structure of the photon
 is twofold.
 Firstly the investigations serve as tests of QED to
 ${\mathcal{O}}(\aem^4)$, but secondly, and also very important,
 the investigations are used to refine the experimentalists tools in a real
 but clean experimental environment to investigate the possibilities of 
 extracting similar information from the much more complex hadronic final
 state.
%
%
\section{The photon structure function \ftqed}
\begin{figure}[htb]
\begin{center}
{\includegraphics[width=1.0\linewidth]{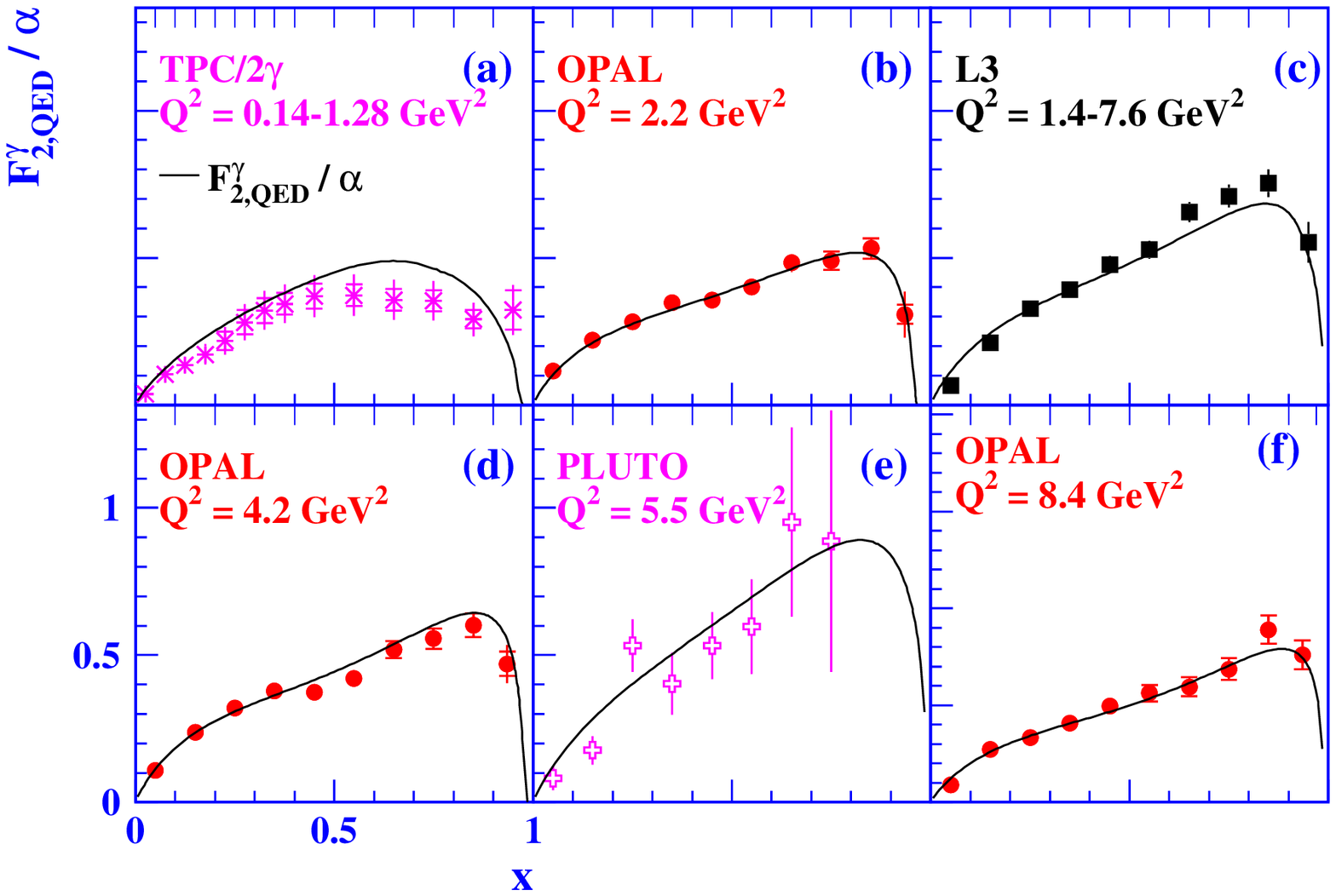}}
\mbox{}\vspace{-1.3cm}
{\includegraphics[width=1.0\linewidth]{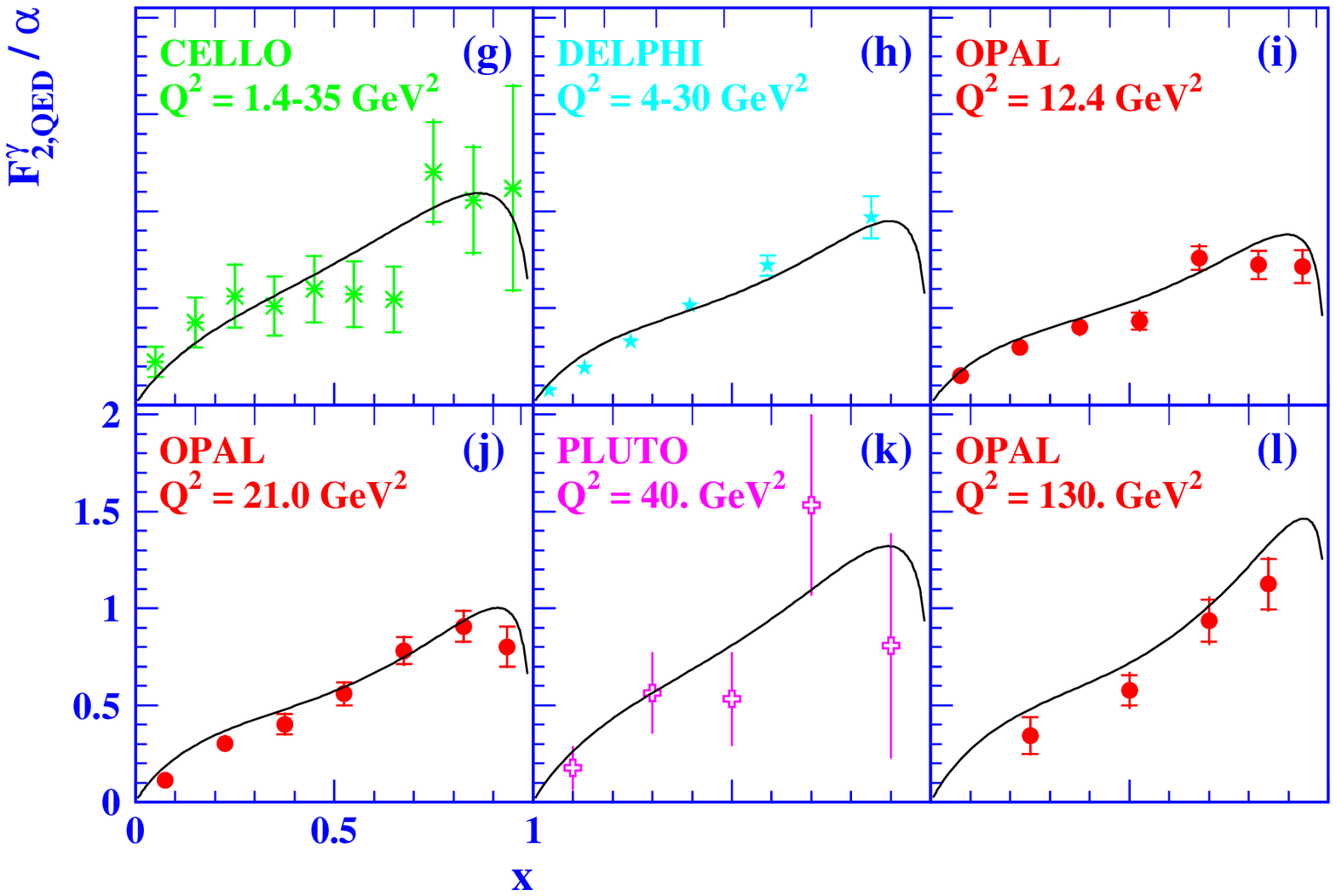}}
\caption{\label{fig01}
         The world summary on \ftqed\ measurements.
         The quoted errors for (h) are statistical only.
        }
\end{center}
\end{figure}

\begin{figure}[htb]\unitlength 1pt
\begin{center}
{\includegraphics[width=1.05\linewidth]{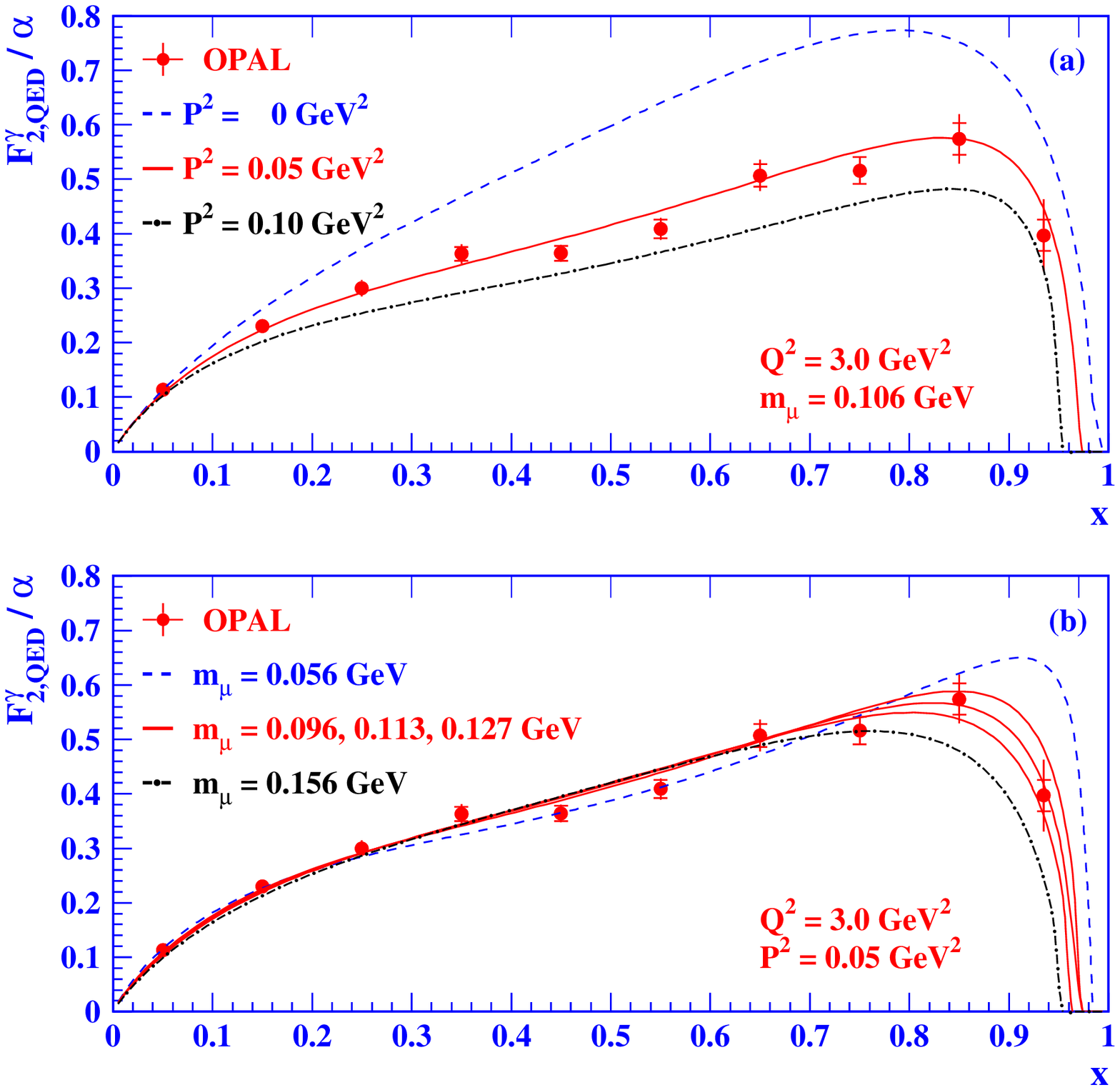}}
\caption{\label{fig02}
         The dependence of \ftqed\ on \psq\ and on the mass of the muon,
         $m_\mu$.
        }
\end{center}
\end{figure}

 The structure function \ftqed\ has been measured using data 
 in the \qsq\ range from about 0.14 up to 400~\gevsq.
 Results were published by the
 CELLO~\cite{CEL-8301},
 DELPHI~\cite{DEL-9601},
 L3~\cite{L3C-9801},
 OPAL~\cite{OPALPR271},
 PLUTO~\cite{PLU-8501}
 and
 TPC/2$\gamma$~\cite{TPC-8401} experiments.
 Special care has to be taken when comparing the experimental results
 to the QED predictions, because slightly different quantities are
 derived by the experiments.
 Some of the experiments express their result
 as an average structure function, $\langle\ftqed(x,\qsq)\rangle$, measured
 within their experimental acceptance in \qsq, whereas the other experiments
 unfold their result as a structure function for an average
 \qsq\ value, $\ftqed(x,\qzm)$.
 Figure~\ref{fig01} shows the world summary of the \ftqed\ measurements 
 compared either to $\langle\ftqed(x,\qsq)\rangle$, assuming a flat 
 acceptance in \qsq, or to $\ftqed(x,\qzm)$, 
 using the appropriate values for \qsq\ and \qzm\ given by the experiments.
 For the measurements which quote an average virtuality \pzm\ of 
 the quasi-real photon for their dataset,
 this value is chosen in the comparison, otherwise $\psq=0$ is used.
 There is a nice agreement between the data and the QED expectations
 for about three orders of magnitude in \qsq.
 The LEP data are so precise that the effect of the small virtuality
 of the quasi-real photon can clearly be established,
 as shown, for example, in Figure~\ref{fig02} for the most precise data
 from OPAL.
%
%
%
 The data are compared to the QED predictions of 
 \ftqed(x,\qzm,\pzm,$m_\mu$), where either \pzm\ or $m_\mu$ is varied.
 Using this data,
 the mass of the muon is found to be $m_\mu=\Y{0.113}{0.014}{0.017}$~\gev,
 assuming the \pzm\ value predicted by QED.
 Although this is not a very precise measurement of the mass of the muon it
 can serve as an indication on the precision possible for the determination
 of $\Lambda$, if it only were for the pointlike contribution to \ft.
%
%
\section{Azimuthal correlations}
\begin{figure}[htb]\unitlength 1pt
\begin{center}
{\includegraphics[width=1.0\linewidth]{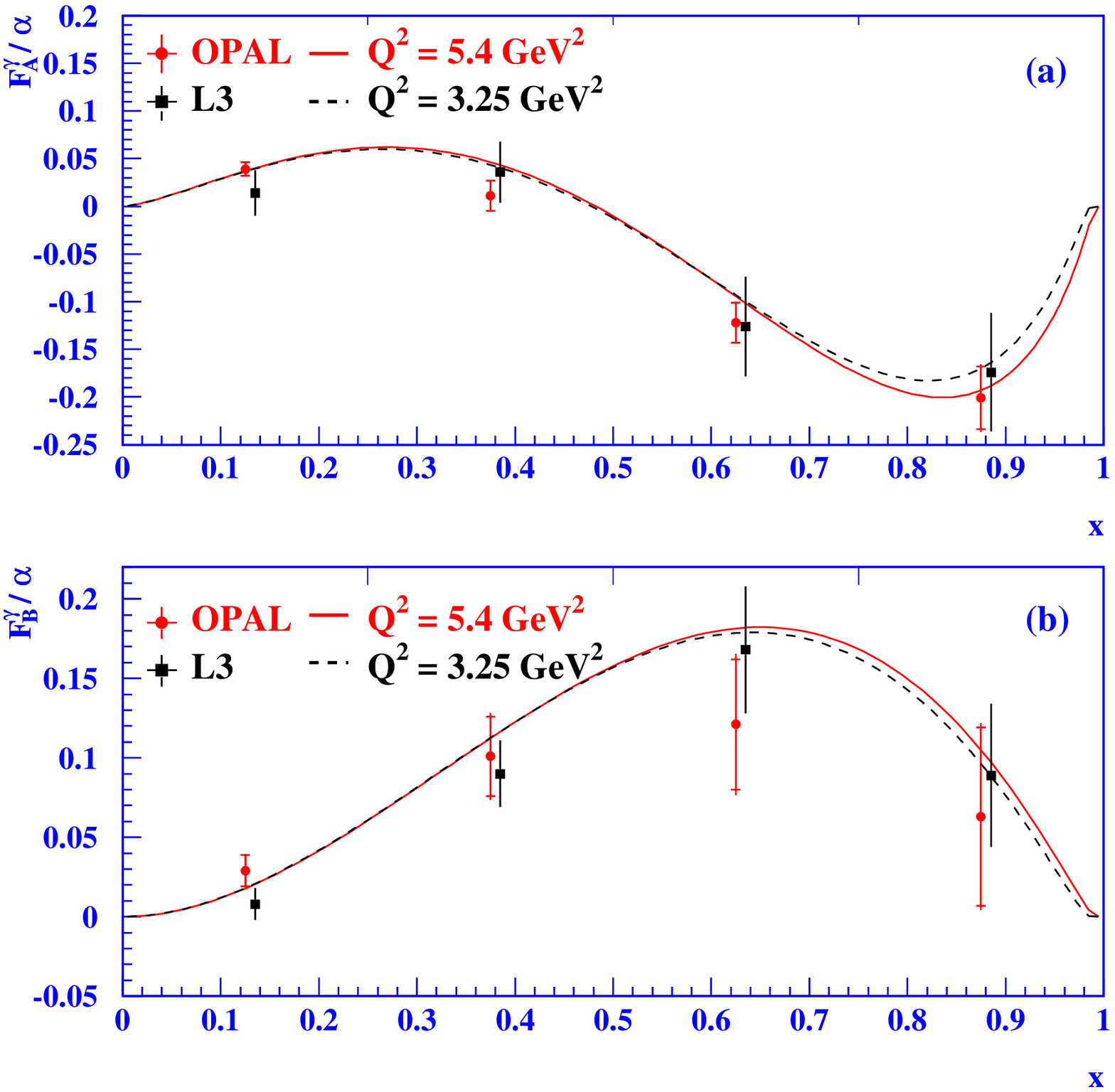}}
\caption{\label{fig03}
         The measurements of \faqed\ and \fbqed\ from L3 and OPAL.
        }
\end{center}
\end{figure}

 The structure functions \faqed\ and \fbqed\ are obtained
 from the measured \ftqed, and a fit to the shape
 of the distribution of the azimuthal angle \az, which is the angle
 between the plane defined by the momentum vectors of the muon pair
 and the plane defined by the momentum vectors of the incoming and the
 deeply inelastically scattered electron.
 For small values of $y$, the \az\ distribution can be written as:
%
 \begin{equation}
 \frac{\der\,N}{\der\,\az} \sim 
 1 - \faoft\cos\az + \frac{1}{2}\fboft\cos 2\az\,.
 \label{eqn:fit}
 \end{equation}
%
 The recent theoretical predictions from Ref.~\cite{SEY-9801} which 
 take into account the important mass corrections up to \omfwq,
 are consistent with the measurements of Refs.~\cite{L3C-9801,OPALPR271},
 Figure~\ref{fig03}.
%
%
 Both \faqed\ and \fbqed\ are found to be significantly different 
 from zero.
 The shape of \fbqed\ cannot be determined very accurately, but it is
 not compatible with a constant.
 The best fit to a constant \fbqed/\aem\ leads to 0.032
 and 0.042 with \chidof\ of 8.9 and 3.1 for the L3 and OPAL
 results respectively.
%
%
\section{Cross-section for highly virtual photons}
 The cross-section for the exchange of two highly virtual photons 
 in the kinematical region under study can schematically
 be written as:
%
 \begin{eqnarray}
 \sigma &\sim&  \stt + \stl + \slt + \sll + \\
        &    &  \frac{1}{2}\ttt\costph - 4\ttl\cosph\, .
\label{eqn:true}
\end{eqnarray}
%
 Here the total cross-sections \stt, \stl, \slt\ and \sll\ and the 
 interference terms \ttt\ and \ttl\ correspond to specific helicity states 
 of the photons (T=transverse and L=longitudinal), and \barph\ is the 
 angle between the electron scattering planes.
%
\begin{figure}[htb]\unitlength 1pt
\begin{center}
{\includegraphics[width=1.0\linewidth]{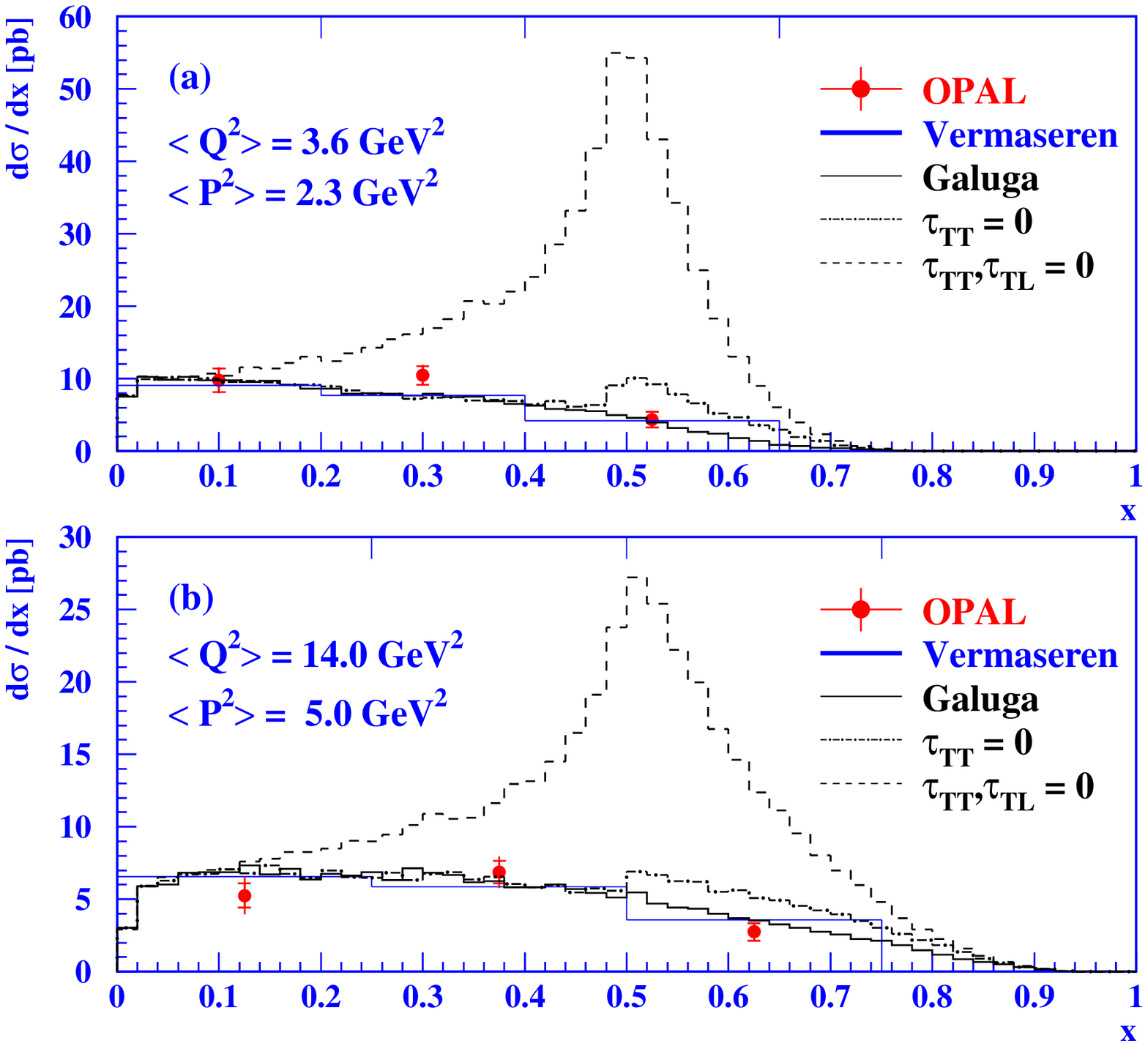}}
\caption{\label{fig04}
         The measurement of the differential cross section \dsigdx\
         for highly virtual photons from OPAL.
        }
\end{center}
\end{figure}
%
 There is good agreement between \dsigdx\ measured by OPAL, 
 Figure~\ref{fig04}, and the QED predictions 
 using the Vermaseren~\cite{VER-8301} and the GALUGA~\cite{SCH-9801}
 Monte Carlo programs, provided all terms of the differential cross-section
 are taken into account.
 However, as apparent from Figure~\ref{fig04}, if either \ttt\ (dot-dash)
 or both \ttt\ and \ttl\ (dash) are neglected in the
 QED prediction as implemented in the GALUGA Monte Carlo, there is a
 clear disagreement between the data and the QED prediction.
 This measurement clearly shows that both terms, \ttt\ and especially \ttl, 
 are present in the data in the kinematical region of the OPAL analysis.
 The contributions to the cross-section are negative and large,
 mainly at $x>0.1$.
 \par
 As the kinematically accessible range in terms of \qsq\ and \psq\
 for the measurement of the QED and the QCD structure
 of the photon is the same, and given the size of the interference
 terms in the leptonic case, special care has to be taken when the
 measurements of the QCD structure are interpreted in terms
 of hadronic structure functions of virtual photons.
%
%
\section{Conclusions}
 QED has been tested to ${\mathcal{O}}(\aem^4)$ using the reaction 
 $\mathrm{e}\mathrm{e} \rightarrow \mathrm{e}\mathrm{e}
 \gamma^{(\star)}\gamma^{\star}
 \rightarrow \mathrm{e}\mathrm{e} \mathrm{\mu}\mathrm{\mu}$, 
 and was found to be in good agreement with
 all experimental results.
 Because the precision of the measurements is limited mainly by the
 statistical error, significant improvements will be made by using
 the full expected statistics of 500~\invpb\ of the LEP2 programme.
 \\
%
%
 {\bf Acknowledgement:}\\
 I wish to thank the organisers of this interesting workshop
 for the fruitful atmosphere they created throughout the meeting. 
%
%

\end{document}